\documentclass[12pt,a4paper,reqno]{amsart}
\usepackage{amsmath,mathtools,cancel,empheq}
    \usepackage{amssymb}
    \usepackage{amsthm}
    \usepackage{amsfonts}
    \usepackage{braket}
    \usepackage{psfrag,color}
    \usepackage{bbold}

\begin{document}
\newcommand{\D}[2]{\frac{\partial#1}{\partial#2}}
\newcommand{\half}{\frac{1}{2}}
\setlength{\oddsidemargin}{-0.5cm}
\setlength{\evensidemargin}{-0.5cm}

\title{Non-particulate Klein-Gordon modes formed by inflation.}
\author{ P. Broadbridge$^1$, K. Deutscher$^2$}
\date{}

\begin{abstract}

In a full solution for a scalar  quantum field coupled to an accelerating isotropic universe, all constituent non-autonomous modes of elementary 
excitation cease to oscillate and become unstable at a discrete sequence of times.  After canonical quantization the time-frozen Hamiltonian has 
eigenstates that can be viewed as neither particles nor oscillatory radiation. Under standard canonical quantization, the Hamiltonian has a natural 
time dependent partitioning into a light component and a dark component. Under equipartition of energy, the energy density of  the dark component 
remains constant. The dark component consists of a finite number of low wave -number repulsive units with time varying force constant. Although these 
unstable modes have no ground state of minimum energy there exist only finitely many of them and so their quantum Hilbert space representation, by the 
Stone-von Neumann theorem, is unique up to unitary equivalence. The remaining infinite number of stable modes still have a unique Fock-Cook 
representation and so overall there is still a preferred physical representation.

\end{abstract}

\maketitle
1 Dept. of Mathematics \and Statistics, La Trobe University, Melbourne VIC 3086, Australia.\\
2. School of Physical Sciences, University of Adelaide SA 5005, Australia.\\

\section{Introduction}

 Over the last decade it has become generally accepted that over intergalactic scales the universe is isotropic, spatially flat, expanding, and 
 accelerating \cite{Ald,Bal,Ber,Pea,Per,Rei,Spe,Teg}.  These assumptions lead to a simple form of the Friedmann equation (e.g.\cite{Car}). Neglecting 
 the radiation component of energy,
$$
\left(\frac{\dot a(t)}{a}\right)^2=\frac{8\pi G}{3}\rho(t)+{\frac{\Lambda c^2}{3}}
$$
where G is the universal gravitational constant, $\rho$ is the  mass density of matter, and $\Lambda$ is the cosmological constant which is assumed to 
be positive to produce an accelerating universe. The energy density of the dark component is $\frac{\Lambda c^4}{8\pi G}$. As time increases and the 
universe expands, the mass density will decrease. Therefore at large times, $a(t)$ will be asymptotic to $e^{\sqrt{\Lambda/3}ct}$.

The classical, generally covariant massless Klein-Gordon field develops unstable modes when it is minimally coupled to the geometry of an exponentially 
accelerating universe \cite{PBPZ,Mijic}. However when unstable classical modes are quantized the Fock representation is no longer physically relevant 
\cite{PBCAH,Paneitz,PBJAMS,PBHadronic}. There are no conserved number operators \cite{Chaiken} and no time-invariant vacuum state from which to build a 
cyclic representation of the quantum field algebra.
It has already been proven \cite{Allen} that no de Sitter-invariant Fock vacuum state exists for a massless minimally coupled field on de Sitter space. 
Allen \cite{Allen} concludes that either (1) the standard  Fock space construction must be abandoned or (2) the vacuum state must break invariance. 
Since breaking invariance is a far less radical idea than the abandonment of the familiar Fock representation, there has been much investigation into 
possible non-invariant vacua for the massless minimally coupled field   \cite{Allen,All&Fol,Kirsten,Polarski}.

 One may instead, consider Allen's first option: the abandonment of the Fock space construction. The  break-down of the Fock representation is entirely 
 due to the unstable modes and as we will show, there are only finitely many of these modes at any time. This means that although these modes have no 
 particle or radiation interpretation via a Fock representation, the Stone-von Neumann theorem guarantees the existence of a preferred physical 
 representation through which we can obtain meaningful calculations and predictions about the system.

In Section 2 the field equations are solved exactly in terms of a discrete expansion of spherical harmonics satisfying appropriate boundary conditions. 
In Section 3 we investigate the energy spectrum. After canonical quantization the Hamiltonian is partitioned into a light and a dark component, the 
independent subsystems of which can be regarded as quantum oscillators with time-dependent frequencies and quantum repulsive units with time-dependent 
force constant akin to Schroer's  ``jelly states" \cite{Schroer}.  As time increases, modes of progressively high wave number cease to be oscillatory 
and join the collection of unbounded repulsive units. The fundamental non-autonomous modes are analysed in two ways. The first is to classify the 
spectrum of the modal Hamiltonian that is frozen in time (Section 3). The second is to directly solve the non-autonomous Schr\"odinger equation that 
evolves from a harmonic oscillator to a repulsive unit (Section 4).
As discussed in Section 5, the repulsive modes are dark in the sense that they have no particle or radiation interpretation and so cannot be detected by 
usual methods that rely on absorption and scattering of particles and oscillatory waves. This phenomenon occurs already at the level of semi-classical 
quantization wherein the scalar field is quantized but the gravitational tensor field is classical.\\  The inflationary period is most simply 
represented by a scalar { inflaton field \cite{Guth} that was coupled to a universe that was expanding exponentially as $a(t)=e^{\sqrt{\Lambda/3}ct}$, 
with $\Lambda$ very large \cite{Starobinsky, Kazanas}. The radius of the observable universe is estimated to be 47 billion light years compared to only 
14 billion light years for the Hubble radius $ct_H$ \cite{NASA}. The favoured explanation is early  large-scale inflation that is evidenced by 
correlations over large space-like separations observed in the cosmic microwave background \cite{PLANCK2}.  The inflationary exponential time scale 
$\sqrt{3/\Lambda c^2}$ must be several orders  of magnitude smaller than the duration of the inflation period, which is supposed to be $10^{-32} s$ }. 
After the sharp big bang, dynamical fields  should have a broad frequency spectrum. As shown in Sections 4 and 5, after many dynamical modes have been 
destabilised at successively higher energy, the critical radial wave number is  proportional to $a(t)$. As is familiar from spherical harmonic 
expansions, the nth principal quantum energy levels  have degeneracy of order $n^2$. These two facts combined, ensure that the dark energy density is 
approximately constant in time during inflation.\\

In Section 6 we discuss the existence of a preferred physical representation of the quantum field algebra of observables, which is conventionally taken 
to be a C$^*$-algebra generated by the Weyl operators \cite{Segal}. Use of a countable basis of exact field solutions in a finite universe has resulted 
in the number of repulsive units remaining finite, although increasing with time. This means that although the repulsive modes have neither a natural 
vacuum representation nor a particle interpretation, their Hilbert space representation is, by the Stone-von Neumann theorem, unique up to unitary 
equivalence. This result is significantly different from that of the approximate continuous Fourier expansion \cite{PBPZ}, since approximation of finite 
space by spatially infinite space, artificially leads to a problematic infinite number of unstable modes. The remaining infinitely many quantum 
oscillator modes still have  a unique Fock-Cook representation, so overall there is still a well defined preferred physical representation.

%------------------------------------------------------------------------------------------------------------

\section{Classical Klein-Gordon field in an accelerating space-time}

{Assuming the geometry of a spatially isotropic expanding universe, the standard metric is
$$ds^2=g_{\mu\nu}dx^\mu dx^\nu=c^2dt^2-a(t)^2\sum_{i=1}^3 dx^idx^i.$$
Here, $x^0=ct$, $x^i$ are to be interpreted as material coordinates that maintain fixed values on matter that is moving with the mean expansion of the 
universe and $t$ is cosmic time that would be measured in such a reference frame.
The generally covariant Klein-Gordon equation is
\begin{equation}\label{gen cov KG}
\nabla^\nu\nabla_\nu\Phi({\bf x},t)+(\frac{m^2c^2}{\hbar^2}+\xi R)\Phi=0
\end{equation}
where $\nabla_\nu$ represents the covariant derivative, }$m$ is the particle rest mass, $\xi$ is the coupling constant for matter wave-gravitational 
field interaction, and {$R=\frac{6}{c^2a^2}(a\ddot a(t)+\dot a^2)$ }is the Ricci Scalar.\\
{ $\nabla^\nu\nabla_\nu$ is the divergence of the gradient or the Laplace-Beltrami operator,
$$\nabla^\nu\nabla_\nu\Phi=|\det g|^{-1/2}\partial_\mu\left[|\det g|^{1/2}g^{\mu\nu}\partial_\nu\Phi\right].$$
When examining the analytic structure of relativistic wave equations, it is convenient  to assume that  $\hbar$ (Planck's constant divided by $2\pi$),  
speed of light $c$ and reference radius $R_0$ have been scaled to unity. The reference radius is the universal radius at a particular reference time. In 
epochs well after inflation, the reference time is usually taken to be the present, labelled here as $t=0$. For studies of  inflation dynamics, the 
reference time may be around $10^{-35} s$ after the big bang. The  length scale $\ell_s$, time scale $t_s$ and mass scale $m_s$ may be chosen so that
$$R_0/\ell_s=1,~~~~c/[\ell_s t_s^{-1}]=1,~~~~\hbar /[m_s \ell_s^2 t_s^{-1}]=1$$
$$\iff \ell_s=R_0,~~~~t_s=R_0/c,~~~~m_s=\hbar/cR_0.$$
After this choice of scaling,  other important constants such as $G$ and $\Lambda$ will be scaled to non-unit dimensionless values:
\begin{eqnarray*}
\hat\Lambda&=&\Lambda\ell_s^2=3\left(\frac{R_0}{ct_H}\right)^2\\
\hat G&=&\frac{G\hbar}{c^3R_0^2}.
\end{eqnarray*}
For convenience, the circumflex will be left off dimensionless variables $\hat x^i=x^i/\ell_s, \hat r=r/\ell_s$ and $\hat t=t/t_s$. However, the 
numerical values of dimensional quantities will be recalled  when necessary.}

Changing to a conformal time coordinate,
 \begin{equation}
 \eta=\int_0^t\frac{1}{a(t_1)}dt_1,
 \end{equation}
 then  the metric conforms to that of Minkowski, $$ds^2=a(t)(d\eta^2-\sum_{i=1}^3 dx^idx^i).$$
 From here on, attention will be focussed on exponential expansion with Hubble time $t_H=\sqrt{3/\hat\Lambda}$:
 \begin{eqnarray}
 a(t)&=&e^{\sqrt{\hat\Lambda /3}t},\\
 \eta&=&\sqrt{3/\hat\Lambda}\left(1-e^{-\sqrt{\hat\Lambda/3}t}\right),\\
 a&=&\frac{\eta_\infty}{\eta_\infty-\eta}~~~~~~\hbox{with}~~ \eta_\infty=\sqrt{\frac {3} {\hat \Lambda}}.
  \end{eqnarray}
Defining $\phi=a\Phi$,  as explained in \cite{PBPZ}, (\ref{gen cov KG}) is expressed as
\begin{equation}\label{KG}
\Box\phi+\mu^2\phi=0
\end{equation}
where
\begin{equation}\label{mu^2}
\mu^2=\left(\frac{3}{\hat\Lambda} {\mathbb m}^2-12\gamma\right)(n_\infty-\eta)^{-2},
\end{equation}
$\gamma=\frac16-\xi$ and $\Box$ is the standard d'Alembertian operator. In these coordinates, the Klein-Gordon equation resembles the standard 
Lorentz-invariant scalar wave equation, allowing one to develop an illustrative quantum field theory using the familiar notation of canonical 
quantization. It is important to note that in terms of dimensional quantities,
{\begin{equation}
\mu^2= \left(\frac{3{\mathbb m}^2c^2}{\Lambda\hbar^2}-\frac{12\gamma}{c^2}\right)(n_\infty-\eta)^{-2}.
\end{equation}
For minimal coupling, $\xi=0$ and $\gamma=1/6$.  In that case the ratio of the mass-dependent term to the mass-independent term within $\mu^2$ is 
$3{\mathbb m}^2c^4/2\hbar^2$. Even for the massive Higgs particle, during the inflation period this ratio is smaller than $10^{-18}$, due to the large 
value of $\Lambda$. }During inflation, particle rest mass has little influence on field dynamics. For illustrative purposes, we will consider ${\mathbb 
m}=0$. Then with minimal coupling $\xi=0$,
\begin{equation}
\mu=i\sqrt 2 (\eta_\infty-\eta)^{-1}=i\frac{\sqrt 2}{t_H}e^{t/t_H}.
\label {musimple}
\end{equation}

 In terms of co-moving (material) space-like coordinates,  the universe is described as the interior of a sphere ${\mathbf x}\cdot{\bold x}<1$. We will 
 also assume the putative boundary condition
 \begin{equation}\label{bc}
 \frac{\partial\phi}{\partial r}=0.
 \end{equation}
 It can be seen by constructing the canonical stress-energy tensor that this condition guarantees zero flux of energy across the material boundary. By 
 separation of variables, the solution to (\ref{KG}) may be expanded in terms of spherical harmonic basis functions

{ \begin{equation}\label{genphi}
F_{nl}(r)Y_l^m(\theta,\varphi)\{a_{nlm}f_{n\ell}(\eta)+b_{nlm}f_{n\ell}^*(\eta)\}
\end{equation}
 where $a_{nlm}$ and $b_{nlm}$ are expansion coefficients. $F_{nl}(r)$ are solutions of
 \begin{equation}\label{Fde}
\frac1F\frac{\partial}{\partial r}\left(r^2\frac{\partial F}{\partial r}\right)+k_{n,\ell}^2r^2=l(l+1),
\end{equation}
wherein $k_{n,\ell}$ are eigenvalues of the radial wave number, to be determined from the boundary conditions.  $f_{n \ell}(\eta)$ satisfies a 
non-autonomous differential equation
% need write the ODE here
 whose solutions are nevertheless well known in terms of Hankel functions or equivalently as Bessel functions,

 $ f_{n \ell}(\eta)=f_{n \ell}^{(1)}(\eta)+i f_{n \ell}^{(2)}(\eta)$, with
 $$
 f_{n \ell}^{(1)}(\eta)=\sqrt{\eta_\infty-\eta}~J\left(\frac12\sqrt{1+48\gamma},k_{n, \ell}(\eta_\infty-\eta)\right)
 $$
 and
 $$
 f_{n \ell}^{(2)}(\eta)=\sqrt{\eta_\infty-\eta}~Y\left(\frac12\sqrt{1+48\gamma},k_{n,\ell}(\eta_\infty-\eta)\right),
 $$
 where $J(\nu,\cdot)$ and $Y(\nu,\cdot)$ are Bessel functions of order $\nu$, of the first and second kind respectively. $F_{n\ell}$ are well-known 
 spherical Bessel functions
\begin{equation}\label{R}
F_{n\ell}(r)=A\frac{J_{l+\frac12}(k_{n,\ell}~r)}{\sqrt{k_{n,\ell} ~r}}+B\frac{Y_{l+\frac12}(k_{n,\ell}~r)}{\sqrt{k_{n,\ell}~r}}.
\end{equation}
Since the Bessel function of the second kind is singular at $r=0$, a finite-valued $\phi$ requires $B=0$. A is absorbed into $a_{nlm}$ and $b_{nlm}$.
Equation (\ref{R}) with the boundary condition (\ref{bc}) determines that in equation (\ref{genphi}), $k_{n,\ell}$ covers the set  of eigenvalues of a 
Sturm-Liouville problem, satisfying
\begin{equation}\label{kvals}
kJ'_{l+\frac12}(k)-\frac12 J_{l+\frac12}(k)=0.
\end{equation}
It follows that the set  of admissible  eigenvalues  $k_{n \ell} $ is the set of zeros  of the derivative spherical Bessel function,
\begin{equation}
j_\ell '(k)=0,
\end{equation}
where $\ell$ is any non-negative integer. $k$ can take one of a discrete set of values being the $s$'th positive zero of $j'_\ell$. In the notation of 
Antosiewicz \cite{Antosiewicz}, $k$ can be $a'_{\ell,s}; \ell=0,1,2,\cdots; s=1,2,\cdots$. It is known that the zeros of $j_\ell'$ and of $j'_{\ell+1}$ 
are interlaced \cite{Liu}:
$$a'_{\ell,1}<a'_{\ell+1,1}<a'_{\ell,2}<a'_{\ell+1,2}<a'_{\ell,3}<\cdots$$
This statement requires $a'_{0,1}$ to be interpreted as 0.
The ordering of zeros of $j'_{\ell}$ is not fully understood. However from Sec. 10.1.58 of \cite{Antosiewicz}, for $s>>\ell$,
\begin{equation}
\label{zeros}
a'_{\ell,s}\sim \beta+\mathcal O(\beta^{-1});~~~\beta=[s+\frac 12(\ell-1)]\pi,
\end{equation}
which is approximately a monotonic function of the integer part $n$ of $\beta /\pi$,$$n=[\beta/\pi]_-=[s+\frac 12(\ell-1)]_-~~.$$
It is convenient to choose
\begin{equation}
k_{n,\ell}=a'_{\ell,[n+1-\frac\ell 2]_-} .
\end{equation}
%$$\phi=a'_{0,1}+\sum_{n=1}^\infty\sum_{l=0}^{2p}
%j_\ell(a'_{\ell,[a']r)Y_l^m(\theta,\varphi)\{a_{nlm}f_{n\ell}(\eta)+b_{nlm}f_{n\ell}^*(\eta)\}$$
The ordering in the following summation is complete, covering all positive zeros of derivative spherical Bessel functions within the factor 
$F_{n\ell}(r)=j_\ell(k_{n,\ell}r)$. It is practically monotonic in $n$ and has the advantage of simple enumeration.
\begin{equation}
\phi=\sum_{n=1}^\infty\sum_{l=0}^{2n}\sum_{m=-\ell}^{\ell}F_{n\ell}(r)Y_\ell^m(\theta,\varphi)\{a_{nlm}f_{n\ell}(\eta)+b_{nlm}f_{n\ell}^*(\eta)\}.
\end{equation}
This expansion is analogous to that which uses the standard ordered basis of energy eigenfunctions  of central force systems \cite{Messiah}. $n$ is 
analogous to the principal quantum number but the $r$-dependence is different because the assumed material surface boundary condition replaces square 
integrability on $\mathbb R^3$. The time dependence is not merely trigonometric since the governing equation is non-autonomous. The basis functions 
$f_{n\ell}$ are chosen to be complex, so that canonical second quantization may proceed just as in the autonomous free field case for which 
$f_{n\ell}(t)$  are complex sinusoids $e^{-i\omega_{n\ell}t}.$
The number of independent basis eigenfunctions with quantum number $n$ is $(2n+1)^2$. Hence, the number of radial wave number eigenvalues that are less 
than or equal to $a'_0,_{n+1}$ is similarly of order $n^2$.}

%-------------------------------------------------------------------------------------------------------------------

\section{Canonical field quantization and unstable modes}
%------------------------------------------------------------------------------------------------------------------

The method of canonical quantization will be the long established approach, similar to that of Pauli \cite{Pauli}. For aspects of coupling with curved 
space-times, the reader is referred to the general introduction by Birrell and Davies \cite{Birrell}. \\

Under canonical quantization, the field $\phi$ becomes a self-adjoint operator of the same form as (\ref{genphi}) but with $a_{nlm}$ and $b_{nlm}$ now 
operators.

{Noting that $(Y_{\ell}^m)^* =(-1)^m(Y_{\ell}^{-m})$}, the requirement that $\phi$ be self adjoint implies  $b_{nlm}^\dagger=(-1)^ma_{nl-m}$, where 
superscript dagger represents the operator adjoint. Therefore the second quantized field is
\begin{equation}\label{phieq}
\phi=
\sum_{n=1}^\infty\sum_{l=0}^{2n}\sum_{m=-l}^l
\frac{J_{l+\frac12}(k_{nl}~r)}{\sqrt{k_{nl}~r}}   Y_l^m(\theta,\varphi)   \{a_{nlm}f_{n\ell}(\eta)+(-1)^ma_{nl-m}^\dagger f_{n\ell}^*(\eta)\}.
\end{equation}

Now the field equation (\ref{KG}) is the Euler-Lagrange equation generated by the Lagrangian density

$$
\mathcal L=\frac12\phi_{,\eta}^2-\frac12(\nabla\phi)^2-\frac12\mu^2\phi^2~,
$$
so the canonical conjugate field variable is
\begin{eqnarray}\label{pi}
\nonumber
\pi(y,\eta)=\frac{\partial \mathcal L}{\partial \dot{\phi}}=\sum_{n=1}^\infty\sum_{l=0}^{2n}\sum_{m=-l}^l
\frac{J_{l+\frac12}(k_{n,l}~r)}{\sqrt{k_{nl}~r}}   Y_l^m(\theta,\varphi)   \{a_{nlm}f'_{n\ell}(\eta)\\
+(-1)^ma_{nl-m}^\dagger f^*_{n\ell}{'}(\eta)\},
\end{eqnarray}
with $\dot\phi=\phi_{,\eta}$.
For Boson fields the field variable and its corresponding canonical conjugate field variable will obey the equal-time commutation relations:
\begin{align}
[\phi({\bf{x}},\eta),\phi({\bf{y}},\eta)]&=0    \label{phicom}  \\
[\pi({\bf{x}},\eta),\pi({\bf{y}},\eta)]&=0      \label{picom}    \\
[\phi({\bf{x}},\eta),\pi({\bf{y}},\eta)]&=i\delta({\bf{x}}-\bf{y})    \label{phipicom},
\end{align}
which taken with (\ref{phieq}) and (\ref{pi}) imply
\begin{align}
[a_{nlm},a_{n'l'm'}]&=0\label{acom}\\
[a_{nlm}^\dagger,a_{n'l'm'}^\dagger]&=0\label{adcom}\\
[a_{nlm},a_{n'l'm'}^\dagger]&=\frac{i}{W_{n\ell}A_{nl}}\delta_{nn'}\delta_{ll'}\delta_{mm'}\label{aadcom}
\end{align}
where $W_{n\ell}$ is the Wronskian $W(f_{n\ell},f_{n\ell}^*)$ and
\begin{equation}\label{Anl}
A_{nl}=\int_0^1\frac{J^2_{l+\frac12}(k_{n,l}~r)}{k_{n,\ell}}rdr{=\frac{2}{k_{n,l}^3}\sum_{p=0}^\infty(\ell+2p)J_{\l+2p+\frac 12}^2(k_{n,l})}~.
\end{equation}
 Since  $A_{nl}$ is a real constant and {$W_{n\ell}=4ik_{n,l}/\pi$ }is a purely imaginary constant, one can normalize $f_{n\ell}$ to
 \begin{equation*}
f_{n\ell}(\eta)= \frac 12 \sqrt{\frac{\pi}{k_{n,\ell}A_{n\ell}}}[f_{n\ell}^{(1)}(\eta)+if_{n\ell}^{(2)}(\eta)]
 \end{equation*}
 so that $W_{n\ell}=\frac{i}{A_{nl}}$ whereby  (\ref{acom}), (\ref{adcom}) and (\ref{aadcom}) are the standard Boson commutation relations and we may 
 interpret $a_{nlm}^\dagger$ and $a_{nlm}$ as Boson creation and annihilation operators respectively.

The canonical Hamiltonian is
\begin{equation}
H =\int_\Omega\frac12\phi_{,\eta}^2+\frac12(\nabla\phi)^2+\frac12\mu^2\phi^2 \,dV~,
\end{equation}
where $\Omega$ is the sectioned 3-space given by $\eta=$constant and $dV$ is the spatial volume element.
To avoid the difficulty of integrating the second term we note that the boundary condition (\ref{bc}) with Green's identity
$$\label{greens}
\int_\Omega\phi\nabla^2\psi+\nabla\phi\cdot\nabla\psi \,dV=\int_{\partial\Omega}\phi\, \hat n \cdot\nabla\psi\, dS
$$
where $\partial\Omega$ is the surface of all space, implies that
\begin{eqnarray}
\nonumber
H&=&\frac12\int_\Omega\phi_{,\eta}^2-\phi\phi_{,\eta\eta}\,dV\\
\nonumber
&=&\frac12 \sum_{n=1}^\infty\sum_{l=0}^{2n}\sum_{m=-l}^l\{W(f_{n\ell}',f_{n\ell}^*)(a_{nl-m}a_{nl-m}^\dagger+a_{nlm}^\dagger a_{nlm})\\
\nonumber
&+&(-1)^mW(f_{n\ell}',f_{n\ell})a_{nl-m}a_{nlm}+(-1)^mW(f_{n\ell}^*{'},f_{n\ell}^*)a_{nlm}^\dagger a_{nl-m}^\dagger \}\\
\nonumber
&=&\sum_{n,l}H_{nl0}+\frac12\sum_{n,\ell,m}A_{nl}\{W(f_{n\ell}',f_{n\ell}^*)(a_{nl-m}a_{nl-m}^\dagger + a_{nlm}a_{nlm}^\dagger\\
\nonumber
&+&a_{nlm}^\dagger a_{nlm}+a_{nl-m}^\dagger a_{nl-m})
+(-1)^mW(f_{n\ell}',f_{n\ell})(a_{nl-m}a_{nlm}+a_{nlm}a_{nl-m})\\
&+&(-1)^mW(f_{n\ell}^*{'},f_{n\ell}^*)(a_{nlm}^\dagger a_{nl-m}^\dagger+a_{nl-m}^\dagger a_{nlm}^\dagger)\}.
\end{eqnarray}
The Hamiltonian is a sum of commuting quadratic Boson terms
\begin{align*}
H_{nl0}&=\frac12  \{W(f_{n\ell}',f_{n\ell}^*)(a_{nl0}a_{nl0}^\dagger +a_{nl0}^\dagger a_{nl0})\\
&+W(f_{n\ell}',f_{n\ell})a_{nl0}a_{nl0}+W(f_{n\ell}^*{'},f_{n\ell}^*)a_{nl0}^\dagger a_{nl0}^\dagger \}\\
\end{align*}
and

\begin{equation}
H_{nlm}=\frac12\left[\begin{array}{cccc}
a_{nlm}^\dagger  &a_{nl-m}^\dagger   & a_{nlm}    &  a_{nl-m}
\end{array}\right]
D_{nlm}\left[\begin{array}{c}
a_{nlm}\\  a_{nl-m}\\    a_{nlm}^\dagger\\      a_{nl-m}^\dagger
\end{array}\right]~\hbox{for~}m\ne 0
\end{equation}
where
$D_{nlm}=A_{nl}\times$
\begin{equation}
\begin{pmatrix}
W(f_{n\ell}',f_{n\ell}^*)                 &0                    &0            &(-1)^mW(f_{n\ell}^*{'},f_{n\ell}^*)\\
        0                &W(f_{n\ell}',f_{n\ell}^*)  &(-1)^mW(f_{n\ell}^*{'},f_{n\ell}^*)              &0\\
        0              &(-1)^mW(f_{n\ell}',f_{n\ell})   &W(f_{n\ell}',f_{n\ell}^*)                  &0\\
        (-1)^mW(f_{n\ell}',f_{n\ell})                 &0                    &0             &W(f_{n\ell}',f_{n\ell}^*)
\end{pmatrix}
\end{equation}

The eigenvalues of $\hat I D_{nlm}$, where $\hat I =\text{diag}[1,1,-1,-1]$, that are invariants of the group of Bogoliubov transformations, are
\begin{equation}\label{eval}
\omega_{nl}=\pm A_{nl}(W(f_{n\ell}',f_{n\ell}^*)^2-W(f_{n\ell}',f_{n\ell})W(f_{n\ell}{'}^*,f_{n\ell}^*))^\frac12.
\end{equation}
These eigenvalues are real when $\mu^2+k_{n,l}^2>0$ and purely imaginary\\ $\omega_{nl}=i\alpha_{nl}$ when $\mu^2+k_{n,l}^2<0$. These are the same 
criteria for real and imaginary frequency found by Mijic \cite{Mijic}. By the theory of equivalence classes of quadratic Hamiltonians \cite{PBPhysica}, 
there exists a Bogoliubov transformation by which the Hamiltonian in terms of new annihilation operators $b_m$ and $b_{-m}$ is
\begin{equation}
H=H_L+H_D
\end{equation}
with
\begin{align}
H_L&=\sum_{\substack{n,l \\ k_{n,l}^2>|\mu^2|}}\frac{\omega_{nl}(\eta)}{2}
\left[  b_{nl0}b_{nl0}^\dagger + b_{nl0}^\dagger b_{nl0}\right]\\
\nonumber & +\frac{\omega_{nl}(\eta)}{2}\sum_{m=1}^{l}\left[b_{nl-m}b_{nl-m}^\dagger + b_{nlm}b_{nlm}^\dagger+b_{nlm}^\dagger b_{nlm}+b_{nl-m}^\dagger 
b_{nl-m}\right]
\end{align}
and
\begin{align}
\nonumber
H_D&=\sum_{\substack{n,l\\k_{n,l}^2<|\mu^2|}}\frac{i\alpha_{n,l}(\eta)}{2}{\left[  b_{nl0}b_{nl0} - b_{nl0}^\dagger b_{nl0}^\dagger\right]}\\
&+\frac{i\alpha_{n,l}(\eta)}{2}\sum_{m=1}^l\left[ b_{nl-m}b_{nl-m}+b_{nlm}b_{nlm}-b_{nlm}^\dagger b_{nlm}^\dagger-b_{nl-m}^\dagger 
b_{nl-m}^\dagger\right].
\label{HD}
\end{align}
We can regard $H_L$ and $H_D$ respectively as a collection of quantum oscillators with time dependent frequency $\omega_k$ and quantum repulsive units 
with  time dependent Hooke coefficient $\alpha_k$. Note that $H_L$ and $H_D$ commute and they each may be further decomposed into a sum of commuting 
Hamiltonian components for independent elementary modes of one degree of freedom. Unlike $H_L$, $H_D$ cannot be transformed to a linear combination of 
number operators since the frequencies are purely imaginary $\omega_k=i\alpha_k$ for all wave numbers $k^2<|\mu^2|$.

The expansion in terms of the countable basis of spherical harmonics results in the number of repulsive modes, although increasing with time, remaining 
finite since for any finite value of $\mu$ there are only finitely many solutions to (\ref{kvals}) with $k^2<|\mu^2|$. The basis modes in (\ref{phieq}) 
have an r-dependent factor
\begin{align*}
\frac{J_{l+1/2}(k_{nl}r)}{\sqrt{k_{nl}r}}~=(-1)^{m+1}\sqrt{\frac{2}{\pi}}\frac{1}{k_{nl}r}\cos(k_{nl}r)+&O([k_{nl}r]^{-3/2})\\
~~&\hbox{for~}l=2p;~p\in Z \\
=(-1)^{m+1}\sqrt{\frac{2}{\pi}}\frac{1}{k_{nl}r}\sin(k_{nl}r)+&O([k_{nl}r]^{-3/2})\\
~~&\hbox{for~}l=2p+1;~p\in Z.
\end{align*}
In this sense,  $k_{nl}$ resembles a wave number. {However, at any time, a field dynamical mode with radial wave number less than $|\mu(\eta)|$ will be 
non-oscillatory, reaching its last extremum near to the time that  $k_{nl}^2-|\mu (\eta)|^2$ becomes negative. At the level of second quantization, that 
dynamical mode is a harmonic oscillator with time dependent squared angular  frequency $\omega^2(t)$ that becomes negative at some finite time. The 
Pauli formulation \cite{Pauli} represents a quantized field as a an infinite system of harmonic oscillators. When the field is minimally coupled to an 
accelerating expanding space-time, there is also a finite number of quantum field modes that may be represented as repulsive units with pure imaginary 
frequency.\\
In the Fock representation of quantum mechanics, a single harmonic oscillator is  represented by the Hamiltonian
$$H=\frac\omega 2[b^\dagger b+bb^\dagger]=\omega[b^\dagger b+1],$$
with $[b,b^\dagger]=1.$  This can be expressed as
\begin{eqnarray*}
H=\frac 12 p^2+\frac 12 \omega^2 q^2\\
q=(2\omega)^{-1/2}[b+b^\dagger],\\
p=-i(\omega /2)^{1/2}[b-b^\dagger],\\
\left[q,p\right]=i.
\end{eqnarray*}
However if the decreasing angular frequency $\omega(\eta)$ becomes zero, the above correspondence between canonical variables $ (q,p)$ and 
creation/ annihilation operators $(b^\dagger,b)$ breaks down.  When the real angular frequency $\omega$ becomes pure imaginary $i\alpha$, the 
self-adjoint time-frozen Hamiltonian $H$ has a continuous spectrum instead of a discrete spectrum and it can no longer be written in terms of a number 
operator $b^\dagger b +1$. Instead, the normal form \cite{PBPhysica} for such a repulsive dynamical mode is $i\frac \alpha 2[bb-b^\dagger b^\dagger]$ as 
in the components of $H_D$ in (\ref{HD}).\\
The spectral properties of the time-frozen Hamiltonian do not give the complete picture of the dynamics of a single component non-autonomous oscillator. 
In the next section, operators $q$ and $p$ will be represented in the usual Schr\"odinger picture as operators on $L^2(\mathbb R)$ with $q: \Psi(x)\to x 
\Psi(x)$ and $p:\Psi(x,\eta)\to -i\partial_x\Psi(x,\eta)$. Thereafter, the relevant non-autonomous Schr\"odinger equation will be solved exactly. Full 
spatial dependence of the field $\phi$ is temporarily disregarded as attention is focused on a  dynamical mode with a single degree of freedom labelled 
by $(nlm)$, with a single pair of canonical variables $(q_{nlm},p_{nlm})$ that are now operators.  In the following section, if $x$ is to be regarded as 
the classical  version of quantum variable $q$, then that variable is associated with the displacement of a single independent dynamical mode of  the 
field $\phi$. However, as in the Pauli construction by Fourier inversion, after each of those separate quantum dynamical modes have been solved exactly, 
they may be recombined to construct the full space-time-dependent field, even though this reconstruction is no longer standard Fourier inversion.}

\section{Non-autonomous quantum evolution: solution with finitely many extrema.}

\subsection{Transforming   to an autonomous Schr\"odinger equation}\quad \\The aim is to solve the non-autonomous Schr\"odinger equation
\begin{equation}
\Psi_{xx}+2i\Psi_\eta-\omega^2(\eta)x^2\Psi=0\label{nonautonomous}
\end{equation}
where
\begin{align*}
\omega^2(\eta)&=k^2-2(\eta_\infty-\eta)^{-2}\\
&=k^2-2\bar\eta^{-2}
\end{align*}
with $\bar\eta=\eta-\eta_\infty<0$, $\eta_\infty=\sqrt{3/\hat\Lambda}$, and $\hat\Lambda$ is the cosmological constant,
where $\eta\in(-\infty,\eta_\infty)$  and $\bar\eta\in(-\infty ,0).$ Note that time evolution, under the non-autonomous equation, is still unitary.  In 
Dirac notation,
$$\frac{d}{d\eta}\ket{\Psi}= -iH(\eta)\ket\Psi ,$$
where $H(\eta)$ is self adjoint and time dependent. If $\ket\chi$ evolves under the same time evolution,
\begin{eqnarray*}
\frac{d}{d\eta}\braket {\Psi(\eta)|\chi(\eta)}&=\braket{-iH\Psi|\chi}+\braket{\Psi|-iH\chi}\\
&=\braket{\Psi|iH\chi}+\braket{\Psi|-iH\chi}=0.
\end{eqnarray*}
By analogous results of Bluman and Kumei on non-autonomous diffusion equations, it is possible to transform to a new function\\ $W=G(x,\eta)\Psi$
satisfying the autonomous Schr\"odinger equation
\begin{equation}
W_{z_1z_1}+2iW_{z_2}-k^2z_1^2W=0,\label{autonomous}
\end{equation}
in terms of transformed variables of the form
$$
z_1(x,\eta),    \qquad      z_2(\eta)
$$
with inverse transformation
$$
x(z_1,z_2),    \qquad       \eta(z_2).
$$

 The above transformation  can be made to approach the identity transformation in the distant past, so that  the non-autonomous equation 
 \eqref{nonautonomous} agrees asymptotically with the autonomous equation \eqref{autonomous}.  Initial conditions for the original pde and the 
 transformed pde will coincide, at a time taken sufficiently far back in the past.
We need to find suitable functions $z_1(x,\eta), z_2(\eta)$ and $G(x,\eta)$.

Now the target autonomous PDE \eqref{autonomous} implies
%\begin{align*}
%&\D{{}^2x}{z_1^2}\left(\D{G}{x}\Psi+G\D{\Psi}{x}\right)+\left(\D{x}{z_1}\right)^2\left(\D{{}^2G}{x^2}\Psi+2\D{G}{x}\D{\Psi}{x}+G\D{{}^2\Psi}{x^2}\right) 
\\
%&+2i\left\{\D{x}{z_2}\left(\D{G}{x}\Psi+G\D{\Psi}{x}\right)+\D{\eta}{z_2}\left(\D{G}{\eta}\Psi+G\D{\Psi}{\eta}\right)\right\}   \\
%&-k^2z_1^2G\Psi=0.
%\end{align*}
%collecting terms%------------equationnnnnnnnnnnnnnnnnnnnnnnnnnnn--------
\begin{align}
\nonumber
&\left(\D{x}{z_1}\right)^2G\Psi_{xx}+\left(\D{{}^2x}{z_1^2}G+2\left(\D{x}{z_1}\right)^2\D{G}{x}+2i\D{x}{z_2}G\right)\Psi_x
+2i\D{\eta}{z_2}G\Psi_\eta  \\
&+\left(\D{{}^2x}{z_1^2}\D{G}{x}+\left(\D{x}{z_1}\right)^2\D{{}^2G}{x^2}+2i\D{x}{z_2}\D{G}{x}+2i\D{\eta}{z_2}\D{G}{\eta}-k^2z_1^2G\right)\Psi
=0.
\label{balance}
\end{align}
Equating coefficients with those of original Schr\"odinger equation \eqref{nonautonomous} gives the following set of PDEs that we must solve, in order 
to find $z_1(x,\eta), z_2(\eta)$ and $G(x,\eta)$.

The ratio of the coefficient of the $\Psi_\eta$ term to the coefficient of the $\psi_{xx}$ term in  \eqref{nonautonomous} must agree with that in 
(\ref{balance}). Therefore,%----------111111111111111111111111111-------------------
\begin{equation*}
 \frac{d\eta}{dz_2}=\left(\D{x}{z_1}\right)^2=T^2,\\
\end{equation*}
where $T$ is an unknown function of $z_2$ only.  Then this gives
\begin{align}
\eta&=\int T^2 dz_2
\quad\text{and}\quad\label{x,eta}
 x=Tz_1+S(z_2)\\
 \implies z_2&=\int \frac{1}{T^2} d\eta \quad\text{and}\quad z_1=\frac{1}{T}(x-S),
\end{align}
where $S$ is some function of $z_2$.
Consequently,
$$
\D{{}^2x}{z_1^2}=0.
$$
With a primed variable denoting a total derivative with respect to $\eta$, we also have
\begin{equation*}
\D{x}{z_2}=T^2\left(\frac{T'}{T}(x-S)+S'\right).
\end{equation*}
Now matching the
$\Psi_x$ terms:%--------------22222222222222222222222---------------------

\begin{align}
\nonumber
&\frac{G_x}{G}=-i\frac{T'}{T}x+iS\frac{T'}{T}-iS', ~~ \hbox{so that}\\
&G(x,\eta)=\exp\left(\frac{-i}{2}\frac{T'}{T}x^2+iS\frac{T'}{T}x-iS'x+F(\eta)\right)\label{G(T,S,F)},
\end{align}
for some function $F$.

This establishes the form of the three functions $z_1, z_2$ and $G$ but we need to specify the three newly introduced functions T, S and F.
%------------TSF------------
From \eqref{G(T,S,F)} we find
$$
\frac{G_{xx}}{G}=-i\frac{T'}{T}+\left(-i\frac{T'}{T}x+iS\frac{T'}{T}-iS'\right)^2
$$
and
\begin{equation}
\frac{G_\eta}{G}=\frac{-i}{2}\left[\frac{T''}{T}-\left(\frac{T'}{T}\right)^2\right]x^2+iS\left[\frac{T''}{T} \label{Geta/G}
-\left(\frac{T'}{T}\right)^2\right]x+iS'\frac{T'}{T}x-iS''x+F'(\eta).
\end{equation}

By comparing the ratios of the $\Psi -$ coefficient to the $\Psi_\eta -$ coefficient in  \eqref{nonautonomous} and in 
(\ref{balance}), it follows that:%----------333333333333333333-----------------------------
\begin{align*}
&\cancel{\D{{}^2x}{z_1^2}\D{G}{x}}+\left(\D{x}{z_1}\right)^2\D{{}^2G}{x^2}+2i\D{x}{z_2}\D{G}{x}+2i\D{\eta}{z_2}\D{G}{\eta}-k^2z_1^2G
=-\left(\D{x}{z_1}\right)^2G\omega^2x^2\\
& \iff T^2\left[-i\frac{T'}{T}+\left(-i\frac{T'}{T}x+iS\frac{T'}{T}-iS'\right)^2\right]G\\
&\qquad+2i\left[T^2\left(\frac{T'}{T}(x-S)+S'\right)\right]\left[-i\frac{T'}{T}x+iS\frac{T'}{T}-iS'\right]G\\
&\qquad+2iT^2\D{G}{\eta}-k^2\left[\frac{1}{T}(x-S)\right]^2G
=-T^2G\omega^2x^2\\
&\iff 2\frac{G_\eta}{G}=i\left[\left(\frac{T'}{T}\right)^2-\frac{k^2}{T^4}+\omega^2\right]x^2
+2i\left[-\left(\frac{T'}{T}\right)^2S+\frac{T'}{T}S'+\frac{k^2}{T^4}S\right]x \\
&\qquad +\frac{T'}{T}+i\left[-2\frac{T'}{T}SS'+\left(\frac{T'}{T}\right)^2S^2+(S')^2-\frac{k^2}{T^4}S^2\right].
\end{align*}
Now compare coefficients with those of the previous expression (\ref{Geta/G}) for ${G_\eta}/{G}$.

From the $x$-coefficient:
\begin{align*}
\cancel{-\left(\frac{T'}{T}\right)^2S}+\cancel{\frac{T'}{T}S'}+\frac{k^2}{T^4}S&=\frac{T''}{T}S 
\cancel{-\left(\frac{T'}{T}\right)^2S}+\cancel{\frac{T'}{T}S'}-S''\\
S''&=-\omega^2S,
\end{align*}
which has general solution
\begin{equation}
S=\sqrt{|\bar\eta|}\left[C_1J_{3/2}(-k\bar\eta)+C_2Y_{3/2}(-k\bar\eta)\right]
\end{equation}
but we will simply choose $S=0$, because this will be most convenient.

From the $x^2$- coefficient:
\begin{align*}
\cancel{\left(\frac{T'}{T}\right)^2}-\frac{k^2}{T^4}+\omega^2&=-\frac{T''}{T}+\cancel{\left(\frac{T'}{T}\right)^2}\\
T''-\frac{k^2}{T^3}+\omega^2T&=0
\end{align*}
which is the Ermakov-Pinney equation with solution
\begin{equation}
C_3 T^2 =k^2 u^2 +u^2\left(C_4+C_3\int \frac{1}{u^2}d\eta\right)^2 \label{S}
\end{equation}
where $u(\eta)$ satisfies
$$
u''+\omega^2 u=0.
$$
We take the simplest non-zero solution for $u$,
\begin{align*}
u&=\sqrt{k|\bar\eta|}J_{3/2}(-k\bar\eta)\\
&=\frac{2}{\pi}(-k\bar\eta)^{-2}[-k\bar\eta\cos(k\bar\eta)-\sin(-k\bar\eta)]^2.
\end{align*}
Then
\begin{align*}
\int\frac{1}{u^2}d\eta &= \frac{1}{k}\int\frac{1}{u^2}d(k\bar\eta)\\
&=-\frac{1}{k}\int\frac{1}{k\bar\eta J^2_{3/2}(-k\bar\eta)}d(k\bar\eta)\\
&=-\frac{\pi}{2k}\frac{k\bar\eta\sin(k\bar\eta)+\cos(k\bar\eta)}{-k\bar\eta\cos(k\bar\eta)+\sin(k\bar\eta)}.
\end{align*}
The singularities due to the zeros in the denominator are cancelled by part of the Bessel term when we multiply by $u$, resulting in $T$ being a well 
behaved function. We would also like to remove the oscillatory behaviour in $T$ due to the trigonometric and Bessel functions, so that the 
transformations to $z_1$ and $z_2$ are one-to-one and invertible. This can be done by
choosing $C_3=2k^2/\pi$ and $C_4=0$, so that

\begin{align*}
\frac{2k^2}{\pi} T^2 &=k^2 u^2 +u^2\left(\frac{2k^2}{\pi}\int \frac{1}{u^2}d\eta\right)^2,\\
            T^2&=(k\bar\eta)^{-2}\left[\left(-k\bar\eta\cos(k\bar\eta)+\sin(k\bar\eta)\right)^2+\left(k\bar\eta\sin(k\bar\eta)+\cos(k\bar\eta)\right)^2\right]\\
                &=1+(k\bar\eta)^{-2},
\end{align*}
which is  1-1 for all times in the domain $\bar\eta\in(-\infty, 0)$.

From the $x^0$-coefficient:

\begin{equation}
2F'(\eta)=\frac{T'}{T}+i\left[-2\frac{T'}{T}SS'+\left(\frac{T'}{T}\right)^2S^2+(S')^2-\frac{k^2}{T^4}S^2\right].
\end{equation}
With the choice, $S=0$,
\begin{align}
2F'(\eta)&=\frac{T'}{T}\\
&=\log(\sqrt{T}).
\end{align}
%---------------------------------------------
We can now substitute these into the transformations.
\begin{align*}
z_2&=\frac{1}{k}\int\frac{1}{T^2}d(k\bar\eta)\\
    &=\frac{1}{k}\int\frac{1}{1+(k\bar\eta)^{-2}}d(k\bar\eta)\\
    &=\frac{1}{k}(k\bar\eta-\arctan(k\bar\eta))+\hbox{const.}\\
\end{align*}

To ensure that the transformation approaches the identity in the distant past (ie as $\eta\rightarrow -\infty$ or $N\rightarrow \infty$) we choose
\begin{equation}
z_2=\bar\eta-\frac{1}{k}\arctan(k\bar\eta)-\frac{\pi}{2k}+\eta_\infty =\eta-\frac{1}{k}\arctan(k[\eta-\eta_\infty])-\frac{\pi}{2k} .
\end{equation}

Next we simply have
\begin{equation}
z_1=\frac{1}{T}(x-S)
\end{equation}

\begin{equation}
\implies z_1=\frac{x}{T}
\end{equation}
which approaches x in the distant past, as desired.

Letting $S=0$ also simplifies G so we have
\begin{align*}
G(x,\eta)&=\sqrt T\exp(-\frac{i}{2}\frac{T'}{T}x^2)\\
                &=(1+(k\bar\eta)^{-2})^{1/4}\exp\left(\frac{i}{2}\frac{kx^2(k\bar\eta)^{-3}}{1+(k\bar\eta)^{-2}}\right),
\end{align*}

which approaches 1 in the distant past, as desired.

\subsection{Harmonic oscillator initially in ground state}\quad \\
 We take W to have to form of a simple harmonic oscillator in the ground state with frequency k (in the distant past, the frequency $\omega(\eta)$ 
 approaches k). $W(z_1,z_2)=(\frac{k}{\pi})^{1/4} e^{-\frac{i}{2}kz_2}e^{-\frac{1}{2}kz_1^2}$,
so
\begin{align*}
\psi(x,\eta)&=\frac{W(z_1,z_2)}{G(x,\eta)}\\
        &=\left(\frac{k}{\pi}\right)^{1/4}(1+(k\bar\eta)^{-2})^{-1/4}\\
        &\times ~\exp\left(\frac{-kx^2}{2}\frac{1-i(k\bar\eta)^{-3}}{1+(k\bar\eta)^{-2}}+\frac i2 \arctan(k\bar\eta)-\frac i2 k\bar\eta+\frac i2[\frac 
        \pi 4-k\eta_\infty]\right).
       \end{align*}

The probability density is
$$
\psi\psi^*=\sqrt{\frac{k}{\pi}}(1+(k\bar\eta)^{-2})^{-1/2}\exp\left(\frac{-kx^2}{1+(k\bar\eta)^{-2}}\right).
$$

Compared to the usual stationary Gaussian probability density of the time-independent oscillator in its ground state, the Gaussian probability density 
spreads as the particle wave number effectively decreases in time, as
$$\bar k=\frac {k(\eta_\infty-\eta)^2}{(\eta_\infty-\eta)^2+1}.$$

\subsection{Superposition of harmonic oscillator ground state and first excited state} \quad \\
 As an example of an oscillating asymmetric state, we take the solution to \eqref{autonomous} to be a superposition of the ground state and the first 
 excited state. We choose the  probability density to be predominantly on one side initially.

 $$W(z_1,z_2)=\cos\Theta\Psi_0(z_1,z_2)+\sin\Theta\Psi_1(z_1,z_2),$$
 which is automatically normalised for any value of $\Theta$. With $\Theta=\pi/4$,

 \begin{align*}
 W(z_1,z_2)&=\frac{1}{\sqrt{2}}(\Psi_0(z_1,z_2)+\Psi_1(z_1,z_2))\\
    &=\frac{1}{\sqrt{2}}\left(\frac{k}{\pi}\right)^\frac{1}{4}\left(e^{-i\frac{k}{2} z_2}+\sqrt{2k}z_1e^{-i\frac{3}{2}kz_2}\right)e^{-\frac{k}{2}z_1^2}
 \end{align*}
 and
 \begin{align*}
 WW*&=\half\sqrt{\frac{k}{\pi}}\left(e^{-i\frac{k}{2} z_2}+\sqrt{2k}z_1e^{-i\frac{3}{2}kz_2}\right)\left(e^{i\frac{k}{2} 
 z_2}+\sqrt{2k}z_1e^{i\frac{3}{2}kz_2}\right)e^{-kz_1^2}\\
    &=\half\sqrt{\frac{k}{\pi}}\left(1+2kz_1^2+2\sqrt{2k}z_1\cos(kz_2)\right)e^{-kz_1^2}.
     \end{align*}

In the original coordinates, the probability density is

\begin{align*}
\Psi(x,\eta)\Psi^*(x,\eta)&=\frac{WW^*}{GG^*}\\
&=\frac{1}{T}WW^*\\
&=\frac{1}{2T}\sqrt{\frac{k}{\pi}}\left(1+2k\frac{x^2}{T^2}+2\sqrt{2k}\frac{x}{T}\cos(kz_2)\right)e^{-\frac{x^2}{T^2}}
\end{align*}

where
\begin{align*}
\cos(kz_2)&=\cos(k\eta-\arctan(k\bar\eta)-\frac{\pi}{2})\\
&=\frac{-k\bar\eta\cos(k\eta)+\sin(k\eta)}{\sqrt{1+(k\bar\eta)^2}}
\end{align*}
which will approach $\sin(k\eta_\infty)$, a constant between -1 and 1 that is related to the exact value of the cosmological constant, since 
$\eta_\infty=\sqrt{\frac{3}{\hat\Lambda}}$.

Now using
\begin{equation*}
\int_0^\infty e^\frac{-x^2}{a} dx=\frac{\sqrt{\pi a}}{2},~~\int_0^\infty xe^\frac{-x^2}{a} dx=\frac a2,~\hbox{and}~\int_0^\infty x^2e^\frac{-x^2}{a} dx 
=\frac{1}{4}\sqrt{\pi}a^\frac{3}{2},
\end{equation*}
the total probability to the right of the origin is

\begin{align*}
\int_0^\infty\Psi\Psi*dx&=\frac{1}{2T}\sqrt{\frac{k}{\pi}}\int_0^\infty 
\left(1+2k\frac{x^2}{T^2}+2\sqrt{2k}\frac{x}{T}\cos(kz_2)\right)e^{-\frac{x^2}{T^2}} dx \\
&=\frac{1}{2}+\frac{1}{\sqrt{2\pi}}\cos(kz_2).
\end{align*}

The probabilistically favoured location oscillates from side to side but eventually the oscillations cease. The extrema in the probability of $x>0$ 
occur at times
$$z_2=\frac{n\pi}{k}\in(-\infty,\eta_\infty-\frac{\pi}{2k}),~~\hbox{with}~n\in{\mathbb Z}.$$
The final extremum occurs at time $z_2=P\frac\pi k$, where P is the largest integer less than $\frac k\pi \eta_\infty-\frac 12.$
The final distribution will depend on the cosmological constant and the limit of $\cos(kz_2)$. In that final distribution, the probability of  $x$ being 
positive is $\frac 12+\frac{1}{\sqrt{2\pi}}\sin(k\eta_\infty).$

\section{Cosmological Implications}
The highest radial wave number for an unstable mode is given by
\begin{equation}
k_{N,l}^2\approx\mu^2=\mathcal O(\eta_\infty-\eta)^{-2}=\mathcal O(a(t)^2)
\end{equation}
for some index $(N,l)$. Combining this with (\ref{zeros}), $N=\mathcal O(a(t))$.
With $n$ fixed, the number  of independent  basis functions is approximately $(2n+1)^2$. Hence the number of independent unstable non-particulate modes, 
is approximately the number of basis functions with $n\le N$, which is $$\sum_{n=1}^N (2n+1)^2=\frac 43 N^3 +\mathcal O(N^2).$$
The number of independent non-oscillatory modes is of order $a(t)^3$. If one assumes equipartition of energy among modes originating from a 
broad-spectrum sharp explosion, then the energy density summed over non-particulate modes will remain approximately constant during the expansion. 
{In more detail, the critical radial wave number is
\begin{equation}
k_{N,l}=|\mu(t)|=\left(\frac{2\hat\Lambda}{3}\right)^{1/2}a(t)\approx N\pi.
\label{kmax}
\end{equation}
The number of independent unstable modes is
\begin{equation}
\sum_{n=1}^N(2n+1)^2=\frac 43 N^3 +6N^2+\frac {17N}{3}.
\end{equation}
If the dimensionless mean energy distributed to each unstable mode is $E$,  then the dark energy density is
\begin{equation}
\rho_{\Lambda}=\frac{\frac 43 N^3 E}{\frac 43 \pi a^3}=2^{-3/2}\pi^{-4}\left(\frac{\hat\Lambda}{3}\right)^{3/2}E.
\end{equation}
This could not account for all dark energy  $\hat\Lambda /8\pi\hat G$ that is consistent with the Friedmann equation unless the energy per unstable mode 
were to be
$$E=\left(\frac 32\right)^{3/2}\pi^3\frac{{\hat\Lambda}^{-1/2}}{\hat G}.$$
$E$ diverges as $\Lambda$ approaches zero, consistent with unstable quantum field modes not being prevalent when accelerated expansion is negligible. 
Conversely, when $\Lambda$ is very large, the dark energy is allocated in small amounts among very many unstable discrete modes, consistent with the 
classical continuum approximation that is inherent in the Friedmann equation.\\
The above presentation includes a number of assumptions that may be rough approximations. The eigenvalue label $n$ is not exactly proportional to 
$k_{n,l}$. The Hamiltonian is not constant but decreasing in time. Similarly, $E(t)$ may be decreasing in time as would be the case at thermal 
equilibrium with radiation at decreasing temperature. Assume that the best-fit power law is
$$k_{n,l}=k_0n^{1/(1+\delta)}=|\mu(t)|=\sqrt 2\left(\frac \Lambda 3\right)^{1/2} a(t)$$. Then the dark energy density will be
\begin{equation}
\rho_\Lambda=\frac {N^3E(t)}{\pi a(t)^3}=E(t)\left(\frac{2\hat\Lambda}{3k_0^2}\right)^{3(1+\delta)/2}\frac{a(t)^{3\delta}}{\pi}.
\end{equation} This agrees with the Friedmann equation if and only if
\begin{equation}
E(t)=\frac 18\left(\frac{3}{2k_0^2}\right)^{3(1+\delta)/2}\hat G^{-1}a(t)^{-3\delta}.
\end{equation}
In the unlikely event that dark energy modes remain in thermal equilibrium with radiation modes whose temperature decreases as $a(t)^{-1}$, $\delta$ 
would be $\frac 13$. In the more likely event that dark energy modes do not freely interact with radiation modes so that they are not in thermal 
equilibrium, but remain near the energy levels at which they became unstable, $\delta$ would be much less than $\frac 13$, consistent with the 
approximation of $k_{N,l}$ being proportional to $N$}.\\

 A scalar inflaton field would have converted a large amount of energy to non-particulate form. This is not the case for the expansion rate that is 
 inferred at the present age of the universe. {The Hubble time $t_H$ is currently estimated to be 14 billion years and the estimated age is 13.7 billion 
 years. Due to early inflation, the radius of the observable universe is much greater than $ct_H$, around 47 billion light years \cite{NASA}. Since the 
 material radius is scaled here to $r=1$, the dimensionless Hubble radius is 14/47. The first positive zero of the spherical Bessel functions  is 
 $k_{1,1}=a'_{1,1}\approx 2.2 $. This mode becomes unstable when $k_{1,1}=|\mu(t)|$. Equivalently from (\ref{musimple}) in dimensionless form,
\begin{eqnarray}
a'_{1,1}=\frac{\sqrt 2}{t_H}e^{t/t_H}\\
\implies \frac{t}{t_H}\approx -0.77,
\end{eqnarray}
which occurred at around 0.23 times the age of the universe which is within the matter-dominated era which lasted until about 0.7 times current age 
\cite{Ryden}.
Here, $t=0$ denotes the present time when $a=1$.  At the present time, the critical radial wave number is $k=\sqrt 2\sqrt{\frac{\Lambda}{3}}a(t)=\sqrt 
2/t_H\approx 4.7.$ If there were already a large number of unstable modes, then we would have $4.7\approx k_{N,l}\approx N\pi$. This implies $N=O(1)$ 
which contradicts the assumption of large $N$.}

Destabilization of the classical field has been examined in detail here only for a massless spin-0 scalar field but we expect that this effect is 
indicative of more general integer-spin fields. We expect that a similar effect will occur for a massless spin-1 field and that Bose quantization of the 
unstable modes will again require a representation with non-particulate eigenstates of the Hamiltonian. The black body spectrum of the cosmic microwave 
background peaks at 1.07$\mu m$. In the scalar field, according to (\ref{kmax}), modes of this wavelength will become repulsive after an exponential 
expansion of approximately 100 $t_H$ duration.  Although at the present time, large-scale cosmogenic field instability is barely incipient and therefore 
has no practical consequence, the unstable non-particulate quantum modes might dominate the energy spectrum in the far-distant future.
%-----------------------------------------------------------------------------------------------------
\section{existence of preferred physical representation}
%-----------------------------------------------------------------------------------------------------
Since the classical field solutions include unstable modes, there is no preferred Fock representation of the Weyl quantum algebra of observables. Even 
if we freeze the form of the Hamiltonian at a particular time, there is no time-invariant vacuum state of zero energy, there is no number operator that 
commutes with the total Hamiltonian \cite{Chaiken} and the energy spectrum of each quantized unstable mode is unbounded below. This is more dramatic 
than the Unruh effect \cite{Unruh} in which observers in different accelerating reference frames ascribe different vacuum states and different 
n-particle states in the same Hilbert space. It is more dramatic than the van Hove effect \cite{Hov} (which in turn is stronger than the Unruh effect), 
in which the strength of interaction requires unitarily inequivalent vacuum representations of the quantum algebra at different times. When a classical 
mode becomes exponentially unstable, the cyclic representation of the quantum algebra based on a time invariant vacuum state is no longer possible 
\cite{PBCAH,Paneitz,PBJAMS}. Each independent unstable mode contributes a continuous spectrum with corresponding generalized eigenfunctions of the 
Hamiltonian that are distributions in the dual of the Hilbert space. These eigenfunctions are neither oscillatory nor particulate in character, but 
related to Schroer's ``jelly" states \cite{Schroer}.  \\

Despite these technical difficulties, there are only finitely many unstable modes at any given time, so that for the unstable subsystem, the Stone-von 
Neumann theorem will guarantee an irreducible separable quantum Hilbert space representation that is unique up to unitary equivalence.
Since the remaining countably infinite set of stable modes can be represented in terms of simultaneous eigenstates of number operators and of the 
Hamiltonian, they can be interpreted as absorption of any number of quanta above the vacuum level. The usual preferred Fock-Cook representation 
\cite{Fock,Cook,Segal} of the stable field component can be extended in a uniquely defined way simply by taking a direct product with the representation 
of the unstable component that has finite degrees of freedom and whose Hilbert space representation is uniquely determined by the Stone-von Neumann 
theorem \cite{Stone,Neumann}.

\section{conclusion}
The effectively massless scalar field minimally coupled to a spatially flat but rapidly expanding FLRW universe with positive cosmological constant, is 
an example of an exactly solvable quantum field theory. It has been shown how each of the contributing non-autonomous quantum dynamical modes may be 
solved exactly even when they evolve from being attractive to repulsive. Like all well-known successful 20th Century quantum field theories, the model 
is semi-classical in the sense that the primary fields are operators but the space-time background is classical. The  pseudo-Riemannian space-time is 
associated with a classical gravitational field. Because of the exponential expansion of the universe, the fundamental dynamical modes of the system 
become unstable at progressively shorter material wavelengths at progressively later times. Much of the energy of the inflaton field would have been 
converted to a non-oscillatory form. Apart from the inflation period, the current universe would admit a relatively small number of unstable 
Klein-Gordon modes. The simple model with minimal coupling already demonstrates the influence of unstable modes on the quantized energy spectrum. A 
single unstable mode contributes a continuous spectrum of the quantized time-frozen Hamiltonian, associated with eigenstates that can be viewed as 
neither oscillatory radiation nor particles. There is no particle number operator that commutes with the time-frozen Hamiltonian of an unstable mode. 
This non-oscillatory character pertains also when the full non-autonomous system is solved; wave components eventually cease to oscillate in time. The 
destabilised energy eigenstates are dark in the sense that neither particle detectors nor wave detectors are designed to interact with them. The 
appearance of these states follows from nothing more than the standard canonical field quantization procedure that has been used for more than 75~years. 
It requires no additional physical postulates.

 The increasing number of independent unstable modes remains finite, whereas the number of stable modes is countably infinite. This allows one to 
 specify a preferred Hilbert space representation with a Fock representation accounting for all of the stable modes. By considering cosmological effects 
 in semi-classical quantization we have focused on effects at extremely long wavelengths. However it is conceivable that a similar mechanism of dark 
 energy generation at short wavelengths may occur in a quantum gravity theory.

\section{Acknowledgement}
The authors are grateful to David Colton (University of Delaware) and Patrick Zulkowski (University of California Berkeley) for  advice on the analysis. 
Sarah Becirevic assisted with a study of  non-autonomous Schr\"dinger equations. The second author gratefully acknowledges support of the Australian 
Research Council for project DP160101366.

\section{Appendix A: Separation of variables}
From the non-autonomous Klein-Gordon equation (\ref{KG}),
\begin{equation*}
\frac{\partial^2\phi}{\partial\eta^2}-\nabla^2\phi-12\gamma(\eta_\infty-\eta)^{-2}\phi=0.
\end{equation*}
In spherical polar coordinates $(r,\theta,\varphi)$,
\begin{eqnarray*}
\nonumber
\frac{\partial^2\phi}{\partial\eta^2}-\frac{\partial^2\phi}{\partial r^2}-\frac 2r\frac{\partial\phi}{\partial r}
-\frac{1}{r^2\sin\theta}\frac{\partial}{\partial\theta}\left[\sin\theta\frac{\partial\phi}{\partial\theta}\right]\\
-\frac{1}{r^2\sin^2\theta}\frac{\partial^2\phi}{\partial\varphi^2}-12\gamma(\eta_\infty-\eta)^{-2}\phi=0.
\end{eqnarray*}
Consider a separated solution $\phi=F(r)L(\theta)Z(\varphi)f(\eta)$ satisfying
\begin{eqnarray*}
\nonumber
\frac{f''(\eta)}{f}r^2\sin^2\theta-12\gamma r^2\sin^2\theta(\eta_\infty-\eta)^{-2}-\frac{F''(r)}{F}r^2\sin^2\theta\\
\nonumber
-\frac 2r\frac{F'(r)}{F}r^2\sin^2\theta-\frac{\sin\theta}{L(\theta)}\frac{d}{d\theta}[\sin(\theta) L'(\theta)]\\
=\frac{Z''(\varphi)}{Z}=-m^2~~~~\hbox{(separation constant).}
\end{eqnarray*}
In order for $Z(\varphi)$ to be $2\pi$-periodic, $m$ can be any integer; consequently $Z(\varphi)=e^{i m\varphi}$. Now separating the independent 
variable $\theta$,
\begin{eqnarray*}
\nonumber
&\frac{-F''(r)}{F}r^2- 2r\frac{F'(r)}{F}+\frac{f''(\eta)}{f}r^2-12\gamma r^2(\eta_\infty-\eta)^{-2}=\\
&\frac{1}{\sin(\theta)L(\theta)}\frac{d}{d\theta}[\sin(\theta)L'(\theta)]-\frac{m^2}{\sin^2\theta}=-\ell(\ell+1)~~~\hbox{(separation constant).}
\end{eqnarray*}
The separated equation for $L(\theta)$ is equivalent to Legendre's equation
$$(1-x^2)\frac{d^2L}{dx^2}-2x\frac{dL}{dx}-\frac{m^2L}{1-x^2}+\ell(\ell+1)L=0,$$
where $x=\cos(\theta).$ $L$ is finite for all $\theta$ if and only if $\ell$ is a non-negative integer. Then $L(\theta)=P_\ell^m(\cos\theta)$, a 
Legendre function of $z$. \\
Separating the remaining independent variables $\eta$ and $r$,
\begin{eqnarray*}
& \frac{F''(r)}{F}&+\frac 2r\frac{F'(r)}{F}-\frac{\ell(\ell+1)}{r^2}\\
=&\frac{f''(\eta)}{f}&-12\gamma(\eta_\infty-\eta)^{-2}=-k^2~~~\hbox{(separation constant).}
\end{eqnarray*}
The separated equation for $f(\eta)$ is equivalent to
$$(\eta_\infty-\eta)^2g''(\eta)-(\eta_\infty-\eta)g'(\eta)+[-(12\gamma+\frac 14)+k^2(\eta_\infty-\eta)^2]g=0,$$
where $g(\eta)=(\eta_\infty-\eta)^{-1/2}f(\eta)$.
Hence the solutions $f(\eta)$ are linear combinations of
$$(\eta_\infty-\eta)^{1/2}J_{\sqrt{12\gamma+1/4}}(k[\eta_\infty-\eta])~~~\hbox{and}~~(\eta_\infty-\eta)^{1/2}Y_{\sqrt{12\gamma+1/4}}(k[\eta_\infty-\eta]).$$
Finally, $F(r)$ satisfies
$$r^2F''(r)+2rF'(r)-\ell(\ell+1)F+k^2r^2F=0.$$
$G(r)=r^{1/2}F(r)$ satisfies Bessel's equation, resulting in the solutions for $F$ being spherical Bessel functions 
$F(r)=j_\ell(kr)=J_{\ell+1/2}(kr)/\sqrt{kr}.$
This neglects the second kind $y_\ell (kr)$ that is undefined at $r=0$.

\end{document}